\documentclass[sigconf]{acmart}
\acmSubmissionID{715}

\usepackage{amsmath}
\usepackage{threeparttable}
\usepackage{multirow}
\usepackage{graphicx}
\usepackage{subfigure}
\usepackage{algorithmic}
\usepackage{algorithm}

\newtheorem{Def}{Definition}

\AtBeginDocument{%
}

\copyrightyear{2023} 
\acmYear{2023} 
\setcopyright{acmlicensed}\acmConference[MM '23]{Proceedings of the 31st
ACM International Conference on Multimedia}{October 29-November 3,
2023}{Ottawa, ON, Canada}
\acmBooktitle{Proceedings of the 31st ACM International Conference on
Multimedia (MM '23), October 29-November 3, 2023, Ottawa, ON, Canada}
\acmPrice{15.00}
\acmDOI{10.1145/3581783.3611853}
\acmISBN{979-8-4007-0108-5/23/10}

\begin{document}

\title{TMac: Temporal Multi-Modal Graph Learning for Acoustic Event Classification}

\author{Meng Liu}
\authornote{Both authors contributed equally to this research.}
\author{Ke Liang}
\authornotemark[1]
\affiliation{%
    \institution{National University of Defense Technology}
        \city{}
    \country{}
}
\email{mengliuedu@163.com}
\email{liangke200694@126.com}

\author{Dayu Hu}
\affiliation{%
    \institution{National University of Defense Technology}
        \city{}
    \country{}
}
\email{hzauhdy@gmail.com}

\author{Hao Yu}
\affiliation{%
    \institution{National University of Defense Technology}
            \city{}
    \country{}
}
\email{csyuhao@gmail.com}

\author{Yue Liu}
\affiliation{%
    \institution{National University of Defense Technology}
        \city{}
    \country{}
}
\email{yueliu19990731@163.com}

\author{Lingyuan Meng}
\affiliation{%
    \institution{National University of Defense Technology}
        \city{}
    \country{}
}
\email{mly\_edu@163.com}

\author{Wenxuan Tu}
\affiliation{%
    \institution{National University of Defense Technology}
        \city{}
    \country{}
}
\email{wenxuantu@163.com}

\author{Sihang Zhou}
\affiliation{%
    \institution{National University of Defense Technology}
        \city{}
    \country{}
}
\email{sihangjoe@gmail.com}

\author{Xinwang Liu}
\authornote{Corresponding author.}
\affiliation{%
    \institution{National University of Defense Technology}
        \city{}
    \country{}
}
\email{xinwangliu@nudt.edu.cn}

\renewcommand{\shortauthors}{Meng Liu et al.}

\begin{abstract}
    Audiovisual data is everywhere in this digital age, which raises higher requirements for the deep learning models developed on them. To well handle the information of the multi-modal data is the key to a better audiovisual modal. We observe that these audiovisual data naturally have temporal attributes, such as the time information for each frame in the video. More concretely, such data is inherently multi-modal according to both audio and visual cues, which proceed in a strict chronological order. It indicates that temporal information is important in multi-modal acoustic event modeling for both intra- and inter-modal. However, existing methods deal with each modal feature independently and simply fuse them together, which neglects the mining of temporal relation and thus leads to sub-optimal performance. With this motivation, we propose a Temporal Multi-modal graph learning method for Acoustic event Classification, called TMac, by modeling such temporal information via graph learning techniques. In particular, we construct a temporal graph for each acoustic event, dividing its audio data and video data into multiple segments. Each segment can be considered as a node, and the temporal relationships between nodes can be considered as timestamps on their edges. In this case, we can smoothly capture the dynamic information in intra-modal and inter-modal. Several experiments are conducted to demonstrate TMac outperforms other SOTA models in performance. Our code is available at \url{https://github.com/MGitHubL/TMac}.
\end{abstract}

\ccsdesc[500]{Information systems~Multimedia information systems}
\ccsdesc[500]{Information systems~Multimedia and multimodal retrieval}
\ccsdesc[300]{Computing methodologies~Artificial intelligence}
\ccsdesc[100]{Computing methodologies~Computer vision}

\keywords{Acoustic event classification, multi-modal audiovisual learning, temporal graph learning}

\maketitle

\section{Introduction}

Audio is an indispensable part of information expression in the real world, and its appearance is often accompanied by visual information. Consequently, many researchers consider visual information as a valuable complement to audio event understanding and analysis \cite{mcgurk1976hearing, atilgan2018integration}. The underlying rationale for this assertion lies in the fact that auditory data is inherently less discernible than its visual counterpart.

\begin{figure}[t]
    \centering
    \subfigure[Audio and video segments over time.]{
        \includegraphics[width=0.49\textwidth]{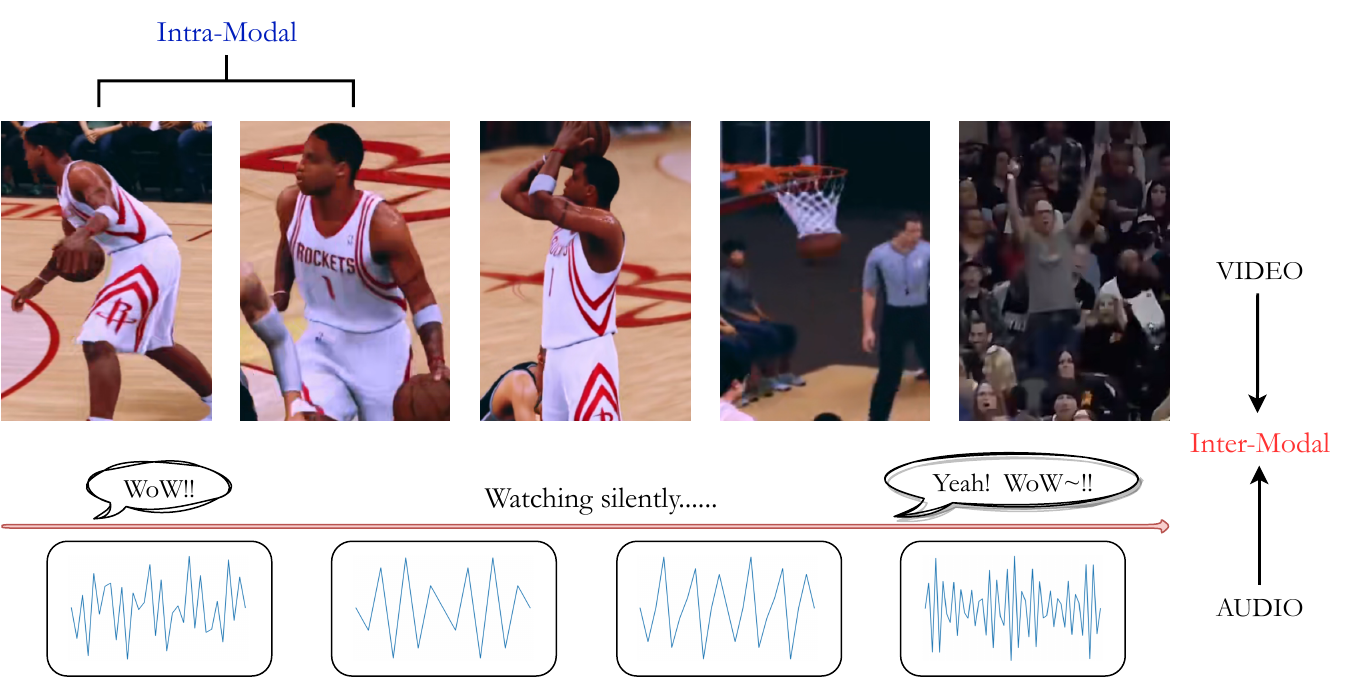}}
    \hspace{0.5in}
    \subfigure[Audiovisual graph construction.]{
        \includegraphics[width=0.49\textwidth]{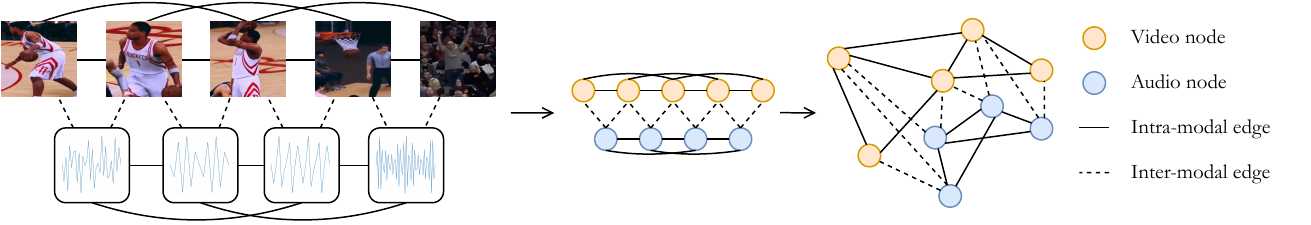}}
    \caption{(a) Visual information as an effective complement to audio events. (b) Such audiovisual data can be well split into multiple segments and constructed as a graph.}
    \label{fig1}
\end{figure}

Imagine the following scenario illustrated in Figure \ref{fig1}, where a basketball athlete is participating in a competitive match. Upon receiving the ball, a handful of spectators couldn't help but exclaim in excitement. Thereafter, the audience fell silent and watched anxiously as the athlete attempted to dunk. Following the successful goal, a multitude of spectators erupted into boisterous cheers. From an auditory standpoint, this can be viewed as an acoustic signal that that fluctuates at first, then falls silent for a while, and eventually changes dramatically. With the arena so noisy, it was difficult for the model to understand a piece of audio that contained a lot of noise. However, when visual signals are added, the logic of audio becomes easier to understand. The model can clearly "observe" the athlete's movements and the spectator's state change. From this perspective, visual data is useful for models to mine important information in audio events.

Acoustic event classification is such a downstream task that has the potential to utilize visual information. Unlike many earlier single-modal audio learning methods, multi-modal audiovisual methods \cite{alayrac2020self, ma2020active} are gaining ground thanks to more established visual research methods that can be quickly fused to audio data. Existing multi-modal methods tend to train visual and auditory views separately and then fuse their representations or construct loss constraints (e.g., contrast loss and mutual information maximization loss) \cite{saeed2021contrastive, shukla2020learning, xihong_ICL_SSL}. In fact, merging audiovisual data into graphs for learning is another potentially valuable route. Although it is relatively rare in the field of acoustics, the graph-based approach has become one of the mainstream approaches in many fields where multi-modal learning is well developed.

It wasn't until recently that a small amount of research \cite{shirian2022visually, shirian2023heterogeneous} introduce the idea of graph construction for modeling audio-visual data, which is exciting. Specifically, by slicing audio and video into segments, each segment can be consider as a node, and the correspondence between them can be consider as edges or relations. Thanks to their contributions, we have gained a fresh perspective. However, there is still some important temporal information that have been overlooked: There is a strict chronological between segments of the same modal, and the corresponding relationship between slices of different modes is precisely anchored by time.

\begin{figure}[t]
    \centering
    \includegraphics[width=0.48\textwidth]{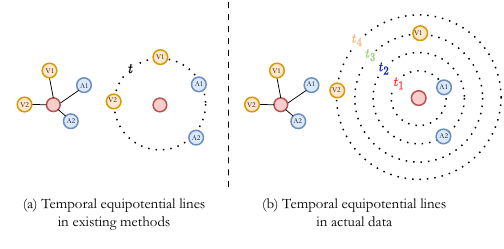}
    \caption{Temporal equipotential lines in different graph constructions. Nodes on the same equipotential line are treated as having equal time status.}
    \label{fig2}
\end{figure}

To illustrate this point more clearly, an example is given in Fig. \ref{fig2}. In this figure, the audiovisual event is constructed as a graph, with each node representing a video or audio segment. Existing graph-based methods consider all nodes (video and audio) to be on the same temporal equipotential line, i.e., there is no difference between the nodes in 1s and the nodes in 100s. Yet these nodes should have been distinguished. In the actual data, nodes with different timestamps are distributed on multiple temporal equipotential lines. For a node,  the neighbor nodes on the closer equipotential line (i.e., the closer appearance time) will have more similar characteristics to it, which is logical. Given the circumstances, it is difficult to overlook temporal information as a crucial attribute.

In fact, there are already well-established techniques to mine temporal information in the graph learning field, i.e., temporal graph neural networks. These temporal graph networks focus on the evolution of node behavior over time, and have shown good results in several fields, such as interest recommendation \cite{xiang2010temporal}, knowledge graph \cite{liang2022reasoning}, smart city \cite{wang2020traffic}, computer network \cite{zhou2023htnet}, etc. Therefore, it is a very natural idea to use temporal graph networks to mine the neglected dynamic information in audiovisual data.

Building upon the aforementioned motivation, we propose the TMac method as a solution for acoustic event classification, which employs temporal multi-modal graph learning. In particular, we introduce the Hawkes process, one of the classical technologies of temporal graph modeling, to measure the dynamic relationship between nodes while aggregating the information intra- and inter- modals. Such process argues that in the data dynamic evolution, the closer the time of two samples, the closer they are. This is also very common in audiovisual data, whereby neighboring pieces of audio or video content exhibit a high degree of similarity and are less likely to undergo significant changes. To verify the validity of considering temporal information, we construct multiple experiments on the widely utilized large benchmark dataset (AudioSet), which demonstrate that TMac outperforms other SOTA methods in performance. Finally, the contributions of this paper are three-fold:

\textbf{Problem}. We point out that audiovisual multi-modal data is well-suited for temporal graph modeling, a concept that has been unexplored in the context of acoustic classification field.

\textbf{Algorithm}. We propose the TMac method to construct the temporal graph structure for audiovisual data, which introduce the Hawkes process to capture the dynamic information of intra-modal and inter-modal.

\textbf{Evaluation}. We compare TMac with multiple SOTA methods with several experiments, which prove the effectiveness and flexibility of TMac.


\section{Related Work}

To further explain our motivation and rationale, we first introduce the signle-modal and multi-modal methods in acoustic event classification task. In addition, we point out that multi-modal learning based on graph has not gained much attention in the field of acoustics, but it is already mature in many fields. Finally, consider the strict chronology in audiovisual data, we describe temporal graph learning methods that can capture dynamic information in the graph effectively.

\subsection{Acoustic Event Classification}

Acoustic event classification is an essential downstream task in audio learning. According to the number of modals used in the data, such audio learning models can be divided into single-modal and multi-modal.

\subsubsection{Single-Modal Audio Learning}

Single-modal learning means that methods only utilize the audio-modal data without any other modal, which is very common in the early stage of audio learning. Initially, random forests (RF), support vector machines (SVM), and Gaussian mixture models (GMM) were commonly utilized as the predominant classifiers \cite{phan2014random, zieger2008acoustic}. Recently, deep learning technologies have garnered the attention of researchers \cite{tang2019improved}, and a series of classical methods such as deep neural networks (DNNs) \cite{gencoglu2014recognition}, convolutional neural networks (CNNs) \cite{zhang2015robust, aytar2016soundnet, han2016acoustic}, and recurrent neural networks (RNNs) \cite{kim2017acoustic}, have begun to be contemplated for sound classification tasks. All these methods only focus on the single audio modal, where ignore the possibility of data enhancement from other modals.

\subsubsection{Multi-Modal Audiovisual Learning}

With the repaid development of multi-modal learning in many areas, researchers began to consider the enhancement of acoustic data by other modal information, thus audiovisual learning was born \cite{alayrac2020self, ma2020active}. For example, cross-modal teacher-student methods \cite{owens2016ambient} leverage the dynamic relations between visual and audio information to acquire effective embeddings. Building on this notion in self-supervised learning \cite{arandjelovic2017look}, several studies have utilized a pretext task that involves predicting whether visual and audio information originate from the same source \cite{arandjelovic2018objects, korbar2018cooperative, morgado2021audio, rouditchenko2020avlnet}. Recent advancements include the use of cross-modality clustering in XDC \cite{alwassel2020self}, and the Evolving Losses approach \cite{piergiovanni2020evolving}, which employs multiple single-modal and multi-modal pretext tasks, demonstrating the ability to learn effective representations in different modals.

Note that despite a lot of work on multi-modal audiovisual learning, few researchers have considered transferring these modalities to graph form for learning, and only recently has a small amount of work emerged. In fact, graph-based approaches have become a common paradigm in many multi-modal learning domains.

\subsection{Graph-based Multi-Modal Learning}

As mentioned above, in areas where multi-modal techniques are well developed, there are many graph-based methods have been proposed to capture cross-modal information.

For instance, Yin et al. \cite{yin2020novel} propose a novel approach where the sentence-image pair is treated as a unified graph, enabling the effective capture of diverse semantic relationships among multimodal semantic units. In MM-GNN \cite{gao2020multi}, images are represented as graphs with three subgraphs, each describing visual, semantic, and digital patterns. The messages passed from one diagram to another are then guided to utilize the context in various ways. Weyssow et al. \cite{weyssow2022better} enhance existing pre-trained code language models by learning jointly with concept graph-based graph neural networks. Li et al. \cite{li2022enhancing} propose a graph-based multimodal context modeling approach and applied it to conversational TTS to improve the speaking style of synthetic speech. MMGCN \cite{wei2019mmgcn} is a framework based on multi-modal graph convolutional networks that aims to generate modality-specific representations of users and micro-videos.

These methods adapt different types of graph neural networks to their respective domain data and have become a widely applicable phenomenon, so it is reasonable to use graph learning method on audiovisual data. On this basis, we repeatedly point out that there are few other types of data that lend themselves as naturally to temporal graph-based learning as audiovisual data.

\subsection{Temporal Graph Learning}

Graph neural networks (GNNs) are widely available in the real world \cite{LiangTKDE, liuyue_SCGC, liuyue_survey} and are attracting the attention of researchers \cite{liuyue_DCRN, MGCN, CCGC, Mo_ICML_2023}. By treating samples as nodes and relationships between samples as edges, GNNs can easily capture the underlying relationships and rules between samples through message propagation mechanisms, which are suitable to various types of graphs \cite{chen2023generalizing, li2023attribute, liang2023message, liang2023knowledge, jin2023estimating, liang2023structure}. GNNs have gained significant popularity and are widely employed in various real-world applications, including recommendation \cite{2023wuAdversarial}, community discovery \cite{liuyue_Dink_net, jin2023predicting}, fake news detection \cite{jin2022towards, yang2022reinforcement}, multi-view clustering \cite{ZJPACMMM, jin2023deep,  10.1145/3503161.3547864, wen2023unpaired}, bioinformatics \cite{hu2023scdfc}, hyper-graph analysis \cite{2023wuhypergraph}, image processing \cite{Jin_2021_ICCV, jin2022unsupervised}, etc, because they can find the relationship between samples in changing and multivariate data \cite{GCC-LDA, wan2023autoweighted,  jin2022prototypical}.

As an important technology in GNNs, temporal graph learning considers the dynamic evolution in the graph structure and node features. Temporal graph methods usually transform the time information as an important weight when constructing the graph neural network model. Some of these methods capture the time evolution from a macro perspective (i.e., discrete-time dynamic graph) \cite{sankar2020dysat}, while others focus on time evolution from a micro perspective (i.e., continuous-time dynamic graph) \cite{gao2021equivalence}. 

For example, CTDNE \cite{qu2020continuous} utilizes the random walk technology with the dynamic bias, which can generate node embeddings from multiple random walks over continuous time. HTNE \cite{zuo2018embedding} models the historical neighborhood sequence to calculate the conditional intensity between nodes over time. EvolveGCN \cite{pareja2020evolvegcn} introduces the CNN to capture the adjacency matrix in each static snapshot and then utilizes the RNN to learn the parameters in CNN over time. TGN \cite{rossi2020temporal} is a temporal graph network that stores the historical events of each node and a novel combination of memory modules and graph-based operators are proposed for more computationally efficiency. Jodie \cite{kumar2019predicting} predicts the future node embeddings to compare the truly learned embeddings, which can encourage the model to focus more on node dynamics. MNCI \cite{liumeng2021inductive} mines both neighborhood influence and community influence and inductively generates node embeddings after each interaction. TREND \cite{wen2022trend} considers static graph neural networks in dynamic graphs which can capture the high-order neighborhood information for node representation learning. TGC \cite{TGC_ML} first considers node clustering task on temporal graphs.

In fact, almost all temporal methods perform downstream tasks on graph data, with little consideration for applications in other scenarios \cite{S2T_ML, AGLI_ML_CIKM, li_add1, li_add2, zhang2023learning}. This means that they still view temporal graphs more as data to mine than as concept that can be applied to other scenarios. In this case, we find a close correlation between temporal graphs and audio-visual data, which may provide new ideas for the wide application of temporal graph learning in different fields.

\section{Methods}

\begin{figure*}[htbp]
    \centering
    \includegraphics[width=1\textwidth]{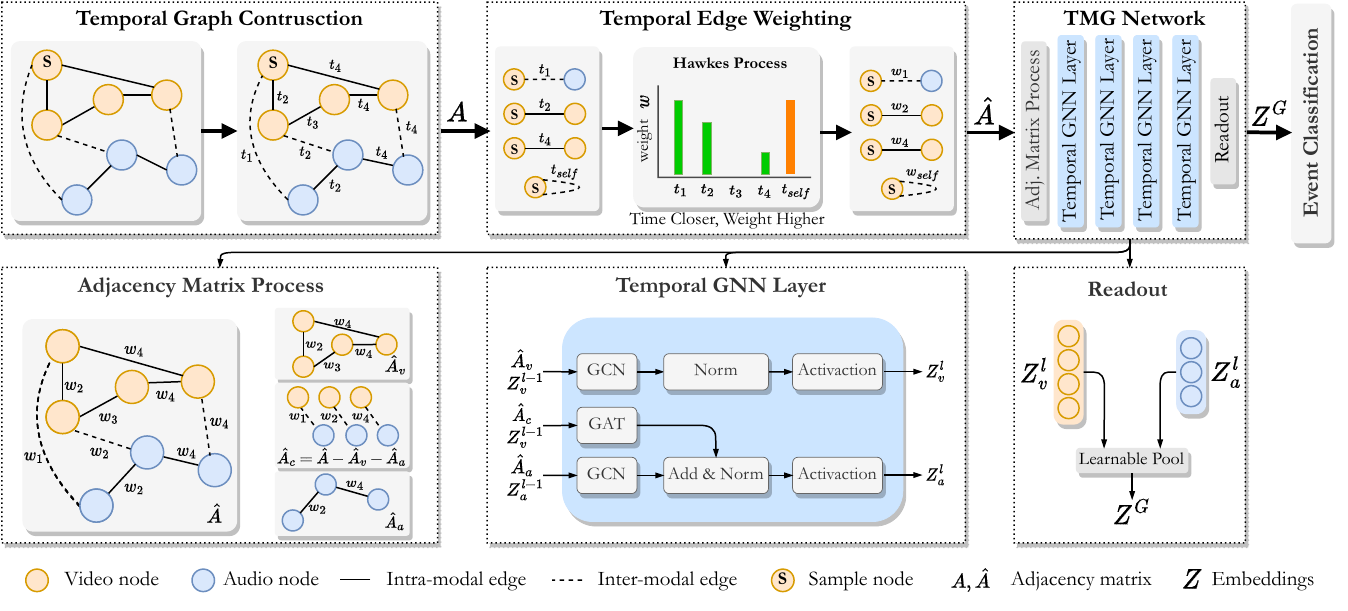}
    \caption{The overall framework of TMac. We first construct an acoustic event as a temporal multi-modal graph, which includes two types of nodes (video and audio nodes) and two types of edges (intra-modal and inter-modal). Before aggregating the neighborhood information of the nodes, we weigh the edges using temporal information. Video nodes and audio nodes are first processed by two GCNs separately, and then information aggregation is completed by GAT. Finally, all these node embeddings are pooled into a whole graph embedding to represent one acoustic event and ultimately used for classification.}
    \label{fig3}
\end{figure*}

\begin{table}[t]
    \centering
    \caption{Notations.}
    \label{Notation}
    \begin{threeparttable}
        \resizebox{0.47\textwidth}{22mm}{
            \begin{tabular}{c c}
                \toprule[2pt]
                Notation& Description\\
                \midrule[1pt]
                $Z^l_a, Z^l_v$& Audio/video node embeddings in the $l$-th layer\\
                $Z^{G_i}$& Graph embedding of the $i$-th event graph\\
                $N_i$& Neighborhood sequence of node $i$\\
                $t_{max}, t_{min}$& The max and min timestamp in one neighborhood\\
                $A_a, A_v, A_c$& Adjacency matrix of audio, video, and cross-modal\\
                $\hat A$& Temporal weighted adjacency matrix\\
                $GCN_a, GCN_v$& Different GCNs for audio and video nodes\\
                $P^a, P^v$& Pooling vectors for audio and video nodes\\
                $M_a, M_v, M_c$& Neighbor number of audio, video, and cross-modal\\
                \bottomrule[2pt]
        \end{tabular}}
    \end{threeparttable}
\end{table}

\subsection{Overall Framework}

We first present the framework of our method TMac. Its inputs are the original audiovisual events, and these events are independent of each other. Our objective is learn a function to classify different acoustic events, thus the outputs are predict labels for them.

In this section, the process of feature extraction and graph construction for the original audiovisual data is given first, and then the model design is given in detail. As shown in Fig. \ref{fig3}, for each event temporal graph, TMac first calculates the temporal weights for each node's neighborhood in it, and then utilizes several GNNs to learn node embeddings. Such embeddings are aggregated into a global representation to represent this event for final classification. Table \ref{Notation} gives the notations.

\subsection{Feature Extraction}

\textbf{Audio Encoding}: To generate features of the audio nodes, each audio clip is partitioned into segments of 960 ms duration, with an overlap of 764 ms. For each audio segment, we compute a log-mel spectrogram using a frame length of 25 ms and 10 ms overlap, 64 mel-spaced frequency bins, and by applying a logarithmic transformation to the magnitude of each bin through short-time Fourier transform. This produces a log-mel spectrogram with dimensions of $96 \times 64$, which is then used as input for the pre-trained VGGish network \cite{hershey2017cnn}. The VGGish network extracts 128-dimensional features for each log-mel spectrogram.

\textbf{Video Encoding}: To extract features of the video nodes, each video is split into non-overlapping chunks of 250 ms duration. To obtain 1024-dimensional features, we pass each segment through an off-the-shelf 3D convolution network, S3D \cite{xie2017rethinking}, which has been trained with self-supervision \cite{han2020self}. It is worth noting that our approach is not limited to these pre-trained embeddings, and can be used with any generic embedding of audio and video.

\subsection{Graph Construction}

Here we discuss the process of constructing acoustic events as graphs. The definition of temporal graphs is given first.

\begin{Def}
    \textbf{Temporal Graphs.}
    A temporal graph can be seen as a graph that contains time evolution. Given a graph $G=(V,E,T,X)$, $V$ and $E$ denote node set and edge set, respectively. $T$ denotes the set of relation timestamps between two nodes, and $X$ denotes the feature set. For any node $i$, there is a neighborhood $N_i$ to store its neighbor nodes and relation timestamp, i.e., $N_i = \{(v_1,t_1),...,(v_m,t_m)\}$.
\end{Def}

In order to model the acoustic events, we construct each event as a temporal multi-modal graph (TMG).

\begin{Def}
    \textbf{Acoustic Events as TMGs.}
    Given an audiovisual event, we uniformly split it into video and audio segments to construct a TMG. The graph consists of two types of nodes (audio and video) and two types of edges (intra-modal and inter-modal). Due to these different types, we can construct three adjacency matrices: video matrix $A_v$, audio matrix $A_a$, and cross-modal matrix $A_c$.
\end{Def}

For the video node $v_i$ and audio node $a_i$, their node embeddings (their feature vectors can be seen as the initial embeddings) can be denoted as $z_i^v$ and $z_i^a$. These nodes have corresponding timestamps from the original segments. For two nodes of the same modality, the edge between them is based on the more posterior timestamp. For two nodes of different modalities, we consider that an edge exists between them only if they are at the same timestamp. Not all edges exist between nodes, and for a node we consider only the nearest $M$ neighbor nodes. $M$ is a hyper-parameter, which we discuss in the experiments.

After constructing each event as a TMG, and then going back to Fig. \ref{fig2}, we can find that it is a natural practice to assign segments from different timestamps to different temporal equipotential lines. This way of graph construction basically adds no additional formalities, and makes the information representation of the graph data more closely match the original audition data.

Next, we will provide details of our model on a single temporal graph. In the actual training, we model these mutually independent events in batches. 

\subsection{Temporal Edge Weighting}

Here we discuss the edge weighting with dynamic information. One of the classic technology in temporal graph modeling is the Hawkes process \cite{hawkes1971point}, which considers the influence of time changes on the closeness of two individuals. For a node $u$'s neighborhood, if neighboring nodes are closer in time to the source node, they tend to be more similar in terms of semantics and features. In this way, when a node receives messages from neighborhood in GNNs, these messages should be weighted by time information.
\begin{equation}
    \hat A = [A_{i,j} W_{i,j}]_{n \times n}
\end{equation}
\begin{equation}
    W_{i,j} = \exp (-\frac{t_{max} - t_i + 1}{t_{max} - t_{min} + 1}), \quad \text{where} j \in N_i
    \label{tem edge weight}
\end{equation}

Here $N_i$ denotes the neighborhood of node $i$. $t_{max}$ and $t_{min}$ denote the max and min timestamp in $N_i$, respectively. This essentially weights the time on each edge so that neighbors with smaller time intervals can receive greater weights. Further, when all three adjacency matrices are weighted by time, $\hat A_v$, $\hat A_a$, and $\hat A_c$ are fed into the GNNs instead of the original matrices.

\subsection{Temporal Multi-Modal Graph Network}

The classical paradigm of GNN is to use the adjacency matrix to aggregate the features or embeddings of nodes through the way of message passing, so that GNN at layer $l$ can combine the $l$-order neighborhood information. The $l$-layer of GNNs can be formulated as follows.
\begin{equation}
    Z^{l} = \text{GNN}(A,Z^{l-1}) = \sigma(A Z^{l-1} W^{l-1}) 
\end{equation}

In this equation, $Z^{l}$ denotes the $l$-layer node embeddings (the initial embeddings are node features), $W^{l-1}$ is a learnable parameter matrix, $\sigma$ denotes the activation function, such as ReLU. Although there are several variants of GNNs, such as GCN, GAT, GraphSAGE, etc., they can all be expressed in the above paradigm.

According to the paradigm, a temporal multi-modal graph network can be regarded as using several GNNs to extract and aggregate different modal information separately, and the adjacency matrix used in this process is weighted by time. 
\begin{equation}
    Z_a^{l} = \text{GCN}_a(\hat A_a, Z_a^{l-1}) + \text{GAT}(\hat A_c, Z_v^{l-1})
    \label{z_a}
\end{equation}
\begin{equation}
    Z_v^{l} = \text{GCN}_v(\hat A_v, Z_v^{l-1})
    \label{z_v}
\end{equation}

Our final objective is acoustic event classification, thus the information of videos need to be aggregated on audio nodes. For the intra-modal information aggregation, we select two GCNs for audio nodes ($\text{GCN}_a$) and video nodes ($\text{GCN}_v$). For the inter-modal information aggregation, we select GAT to propagate video embeddings to audio embeddings.

Note that we first construct a temporal multi-modal graph for each acoustic event, i.e., all these audio and video nodes come from the entire event. After modeling the node information, we need to learn a graph readout function for this event to pool all node embeddings into one final embedding. For the $i$-th event $G_i$, its graph embedding can be calculated as follows.
\begin{equation}
    Z^{G_i} = \text{Readout}(G_i) = \left[\text{P}(Z_a^l) \mid \text{P}(Z_v^l)\right] = Z_a^l P^a + Z_v^l P^v
    \label{readout}
\end{equation}

$\text{P}(x)$ is the pooling function that discards the existing functions (such as MAX, MEAN, and SUM) in favor of constructing a learnable pooling vector $P$. We use a pooling function to generate the graph embedding $Z^{G_i}$ for the $i$-th event by feeding the final $l$-layer node embeddings into it.

\subsection{Loss Function}

The overall temporal multi-modal graph network is trained using focal loss $L$ as follows.
\begin{equation}
    L = - \sum_n(1 - \boldsymbol{y}_n)^{\gamma} \log \boldsymbol{\tilde{y}}_n
    \label{loss}
\end{equation}

Note that the focal loss \cite{lin2017focal} is proposed to solve the category imbalance and the difference in classification difficulty, where $\boldsymbol{\tilde{y}}_n$ denotes predict labels and $\boldsymbol{y}_n$ denote ground-truth labels. The modulating factor $\gamma$ reduces the contribution of easy examples to the overall loss and extends the range in which an example receives low loss. In our implementation, we set $\gamma$ to 2. The procedure of TMac is shown in Algorithm \ref{TMac code}.

\begin{algorithm}[t]
    \caption{TMac procedure}
    \label{TMac code}
    \begin{algorithmic}[1]
        \REQUIRE Acoustic events and corresponding ground-truth labels.\\
        \ENSURE The predict labels of all events.\\
        \STATE {Construct a TMG $G=(V,E,T,X)$ for each event;}
        \STATE {Fetch data by batch and Initialize model parameters;}
        \REPEAT
        \FOR {each $epoch$}
        \FOR {each $batch$}
        \STATE {Calculate temporal edge weights based on Eq. \eqref{tem edge weight};}
        \FOR {each layer in GNNs}
        \STATE {Update video node embeddings based on Eq. \eqref{z_v};} 
        \STATE {Update audio node embeddings based on Eq. \eqref{z_a};}
        \ENDFOR
        \STATE {Readout the graph embedding on Eq. \eqref{readout};}
        \ENDFOR
        \STATE {Optimize the loss function based on Eq. \eqref{loss};} 
        \ENDFOR
        \UNTIL{Convergence} 
    \end{algorithmic}
\end{algorithm}

\section{Experiments}

Here we report the performance and effectiveness of TMac. Firstly, we describe the dataset we used and the baselines we compared. Then we provide model parameter settings and training configurations. In addition to comparisons with multiple methods for node classification experiments, we also conducted ablation study, parametric sensitivity study, and convergence analysis.

In order to introduce each group of experiments more clearly, we put forward several questions and answered them separately according to the experimental results:

\textbf{Q1}: Is temporal information really useful for the acoustic event classification task?

\textbf{Q2}: Is temporal information both intra-modal and inter-modal meaningful?

\textbf{Q3}: Does the model's effect remain stable when faced with different parameter choices?

\textbf{Q4}: Will the extra consideration of time information affects the convergence of TMac?

\subsection{Dataset}
Our experiment is based on a widely used dataset, AudioSet \cite{gemmeke2017audio}, which is also the most representative publicly available data set that is fully suitable for audio-visual data learning. In fact, many methods for acoustic event classification have also conducted relevant experiments on this one dataset only.

Audioset is a manually annotated dataset of audio events from the YouTube platform, created to bridge the gap in data availability between image and audio research. The dataset consists of 10-second segments of YouTube videos, manually tagged to detect the presence of specific audio classes. In our experiment, we selected 33 types of data with high rater confidence scores ({0.7, 1.0}), resulting in a training set of 82,410 audiovisual clips. For a fair comparison with the baseline method, we used the original evaluation set, which contained 85,487 test clips \cite{shirian2022visually}. The dataset was split into three sets for training: a train set (70\%), an evaluation set (10\%), and a test set (20\%).

\subsection{Baselines}

We compare TMac with multiple state-of-the-art baseline methods.

\textbf{Spectrogram-VGG} \cite{simonyan2014very} is a classic method for deep visual representations.

\textbf{DaiNet} \cite{dai2017very} is a network based on 1D convolutions that operates on the raw audio waveform.

\textbf{R(2+1)D} \cite{tran2018closer} is a new spatio-temporal convolutional module based on CNNs.

\textbf{Wav2vec2} \cite{baevski2020wav2vec} masks the speech input in the latent space and solves a contrastive task defined over a quantization of the jointly learned latent representations.

\textbf{Wave-Logmel} \cite{kong2020panns} is a supervised CNN model that takes both the waveform and log mel spectrogram as input.

\textbf{ResNet-1D} \cite{he2016deep} is a fully 1-D ResNet, which uses audio and audio-video inputs, respectively.

\textbf{VATT} \cite{akbari2021vatt} is a self-supervised multimodal transformer that uses a single backbone transformer with shared weights for both audio and video modalities.

\textbf{AST} \cite{gong2021ast} is a self-supervised transformer model trained by masking the input spectra.

\textbf{PaSST-S} \cite{koutini2021efficient} is a transformer-based method that optimizes and regularizes transformers on audio spectrograms.

\textbf{VAED} \cite{shirian2022visually} uses heterogeneous graphs to explicitly capture the relationships between modalities, providing detailed information about the underlying signal.

\textbf{Audio-MAE} \cite{huang2022masked} extends the image-based MAE technique to self-supervised representation learning from audio spectrograms.

\textbf{MaskSpec} \cite{chong2022masked} is a transformer-based method that masks random patches of the input spectrogram and reconstructs the masked regions using an encoder-decoder architecture.

\textbf{SSL graph} \cite{shirian2022self} is a sub-graph-based framework for learning effective audio representations.

\textbf{HGCN} \cite{shirian2023heterogeneous} is a further extension of the heterogeneous graph cross-modal network of VAED.

\subsection{Task and Parameter Settings}

For each video clip in our dataset, we generate a temporal multimodal graph consisting of 40 audio nodes and 100 video nodes, where each node corresponds to a 960 ms audio clip or a 250 ms video clip. To ensure the robustness of our experiments, we repeat each experiment 10 times with different random seeds and report the mean average precision (mAP) and area under the ROC curve (AUC) values.

We initialize our network weights using Xavier initialization and employ the Adam optimizer with a learning rate of 0.005 and a decay rate of 0.1 after 250 iterations, along with 1000 warm-up iterations for all experiments. We select 8 neighbors for each node, including both audio, video, and cross-modal types. Our TMac model consists of 4 layers with a hidden size of 512, where each layer is composed of GCN and GAT. All experiments are implemented using PyTorch on an NVIDIA RTX-3070Ti GPU.

\subsection{Acoustic Event Classification}

\begin{table}[t]
    \centering
    \caption{Acoustic event classification performance.}
    \label{performance}
    \begin{threeparttable}
        \begin{tabular}{c|c c c}
            \toprule[2pt]
            Model (Year)& mAP& AUC& Params\\
            \midrule[0.5pt]
            Spectrogram-VGG& 0.26$\pm$0.01& -& 6M\\
            ResNet-1D-audio& 0.35$\pm$0.01& 0.90$\pm$0.00& 40.4M\\
            ResNet-1D-both& 0.38$\pm$0.03& 0.89$\pm$0.02& 81.2M\\
            DaiNet& 0.25$\pm$0.07& -& 1.8M\\
            R(2+1)D-video& 0.36$\pm$0.00& 0.81$\pm$0.00& 33.4M\\ 
            Wav2vec2-audio& 0.42$\pm$0.02& 0.88$\pm$0.00& 94.4M\\
            Wave-Logmel& 0.43$\pm$0.04& -& 81M\\
            VATT& 0.39$\pm$0.02& -& 87M\\
            AST& 0.44$\pm$0.00& -& 88M\\
            PaSST-S& 0.49$\pm$0.01& 0.90$\pm$0.01& 87M\\
            VAED& \underline{0.50$\pm$0.01}& \underline{0.91$\pm$0.01}& 2.1M\\
            Audio-MAE& 0.47$\pm$0.01& -& 86M\\
            MaskSpec& 0.47$\pm$0.02& -& 86M\\
            SSL graph& 0.42$\pm$0.02& -& 218K\\
            HGCN & 0.44$\pm$0.01& 0.88$\pm$0.01& 42.4M\\
            \midrule[0.5pt]
            TMac& \textbf{0.56$\pm$0.01}& \textbf{0.94$\pm$0.01}& 4.3M\\
            (improv.)& (+12.00\%)& (+3.29\%)& -\\
            \bottomrule[2pt]
        \end{tabular}
    \end{threeparttable}
\end{table}

\textbf{Q1}: Is temporal information really useful for the acoustic event classification task? \textbf{Answer}: Temporal information is useful, and combining it does not result in massive model inflation.

We report the acoustic event classification performance in Table \ref{performance}. According to the experimental results, we can find that: 

(1) TMac outperforms other all methods on the mAP metric. Compared to the model VAED with sub-optimal results, we have improved our results by 12\%.

(2) We also report the AUC performance, and TMac achieves the highest ROC-AUC score of 0.94, indicating that it produces more reliable predictions at various thresholds.

(3) After considering the additional time information, the parameter number of TMac is only 4.5 M. Compared to other models, the small parameter number of TMac means that it is easy to implement.

Based on the above experimental results, we believe that it is valuable to consider time information. In addition, compared to the transformer-based methods (such as VATT and AST), graph-based methods (such as VAED, HGCN, and our TMac) have considerably fewer learnable parameters. This means that graph-based methods are more flexible and easier to deploy in the face of large-scale data.

\subsection{Ablation Study}

\textbf{Q2}: Is temporal information both intra-modal and inter-modal meaningful? \textbf{Answer}: Both intra-modal and inter-modal temporal information are meaningful, with inter-modal information being more meaningful.

In this part, we discuss the effect of these different modules of TMac on the performance, which can be divided into several variants:

(1) \textbf{Non-TMG}: TMac without any temporal information, i.e., VAED. (2) \textbf{Non-intraT}: TMac without intra-modal temporal information. (3) \textbf{Non-interT}: TMac without inter-modal temporal information. (4) \textbf{TMac}: Fully TMac method.

\begin{figure}[t]
    \centering
    \includegraphics[width=0.4\textwidth]{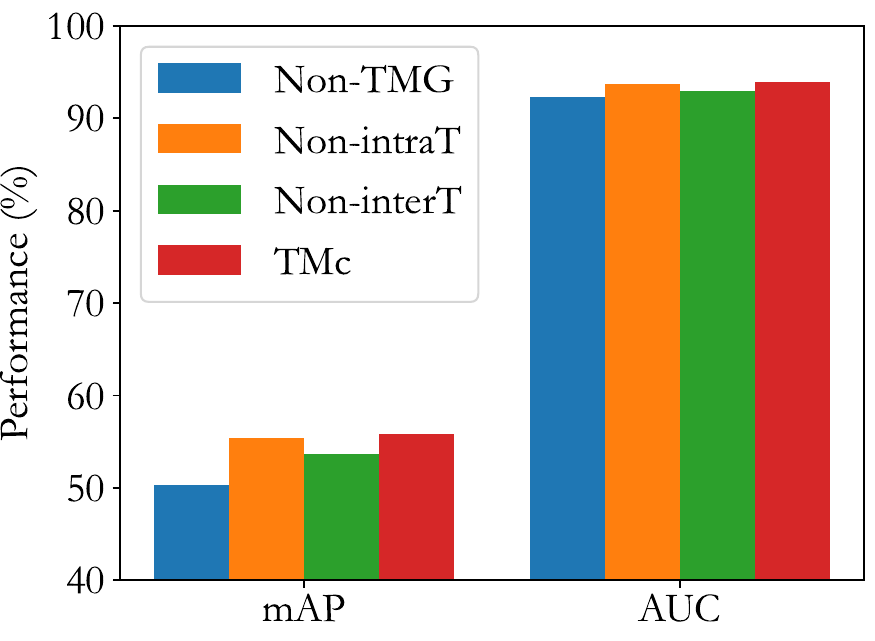}
    \caption{Ablation study on temporal information.}
    \label{ablation}
\end{figure}

According to Fig. \ref{ablation}, we can find that the method without TMGs (Non-TMG) achieves the worst performance, which demonstrates the importance of our temporal multi-modal graph construction. When we discard the intra-modal temporal information (Non-intraT) and inter-modal temporal information (Non-interT) separately, the performance is reduced compared to both fully TMac, but the performance of the Non-interT model is reduced even more. This implies that both intra- and inter-modal temporal information modeling are useful, and that inter-modal temporal relationship are more important.

\begin{figure}[t]
    \centering
    \subfigure[Effect on mAP.]{
        \includegraphics[width=0.23\textwidth]{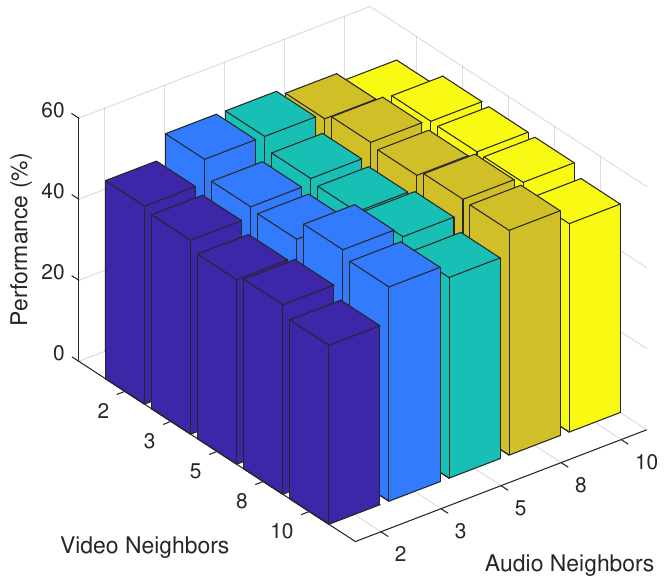}}
    \subfigure[Effect on AUC.]{
        \includegraphics[width=0.23\textwidth]{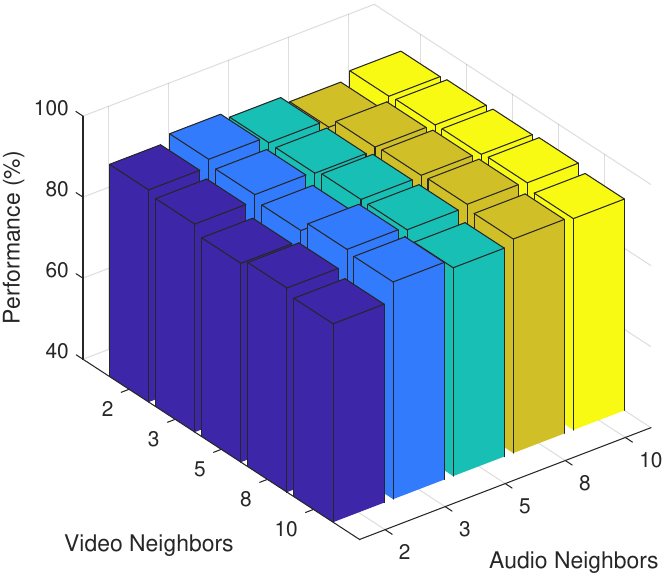}}
    \caption{The effect of different intra-modal neighbor numbers on the mAP and AUC metrics.}
    \label{intra}
\end{figure}

\subsection{Parameter Sensitivity}

\begin{figure}[t]
    \centering
    \subfigure[Effect on mAP.]{
        \includegraphics[width=0.23\textwidth]{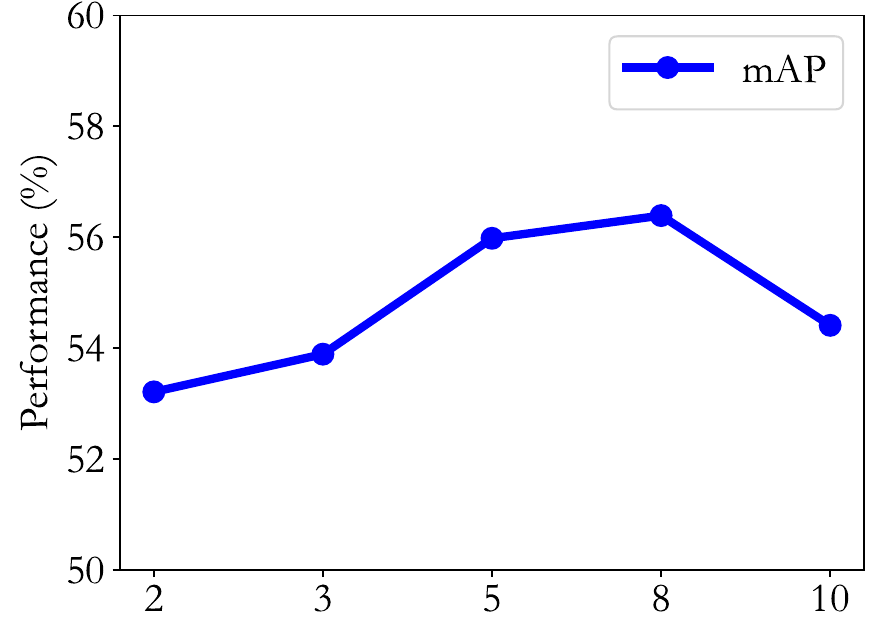}}
    \subfigure[Effect on AUC.]{
        \includegraphics[width=0.23\textwidth]{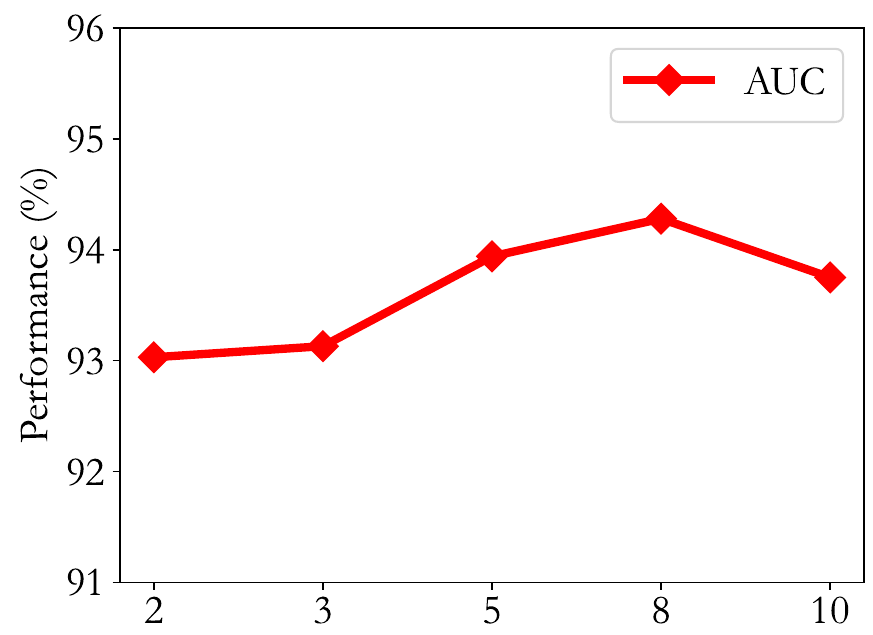}}
    \caption{The effect of different inter-modal neighbor numbers on the mAP and AUC metrics.}
    \label{inter}
\end{figure}

\textbf{Q3}: Does the model's effect remain stable when faced with different parameter choices? \textbf{Answer}: No matter how the super-parameters are combined, the model can maintain a good performance.

We focus on the effect of intra-modal neighbor number $M$, which can be divided into $M_v$ for video-modal, $M_a$ for audio-modal, and $M_c$ for cross-modal respectively, where $M_v=M_a=M_c=8$. To discuss their effects on the performance, we select the values of 2, 3, 5, 8, and 10 for these three parameters respectively to observe the experimental results changes.

\subsubsection{Number of intra-modal neighbors}
For the intra-modal neighbor number, we fix other parameters to observe the performance changes brought about by changing $M_v$ and $M_a$.

As shown in Fig. \ref{intra}, when we select $M_v$ and $M_a$ as 2, they correspond to the lowest model performance. It demonstrates that information propagation from neighbor nodes is efficient in graph structures. Note that this is a reduction in the number of neighbors, not just a reduction in time information. In other words, the performance degradation caused by the reduced number of neighbors actually proves the importance of the neighborhood information. In addition, we can also explain why the performance doesn't look particularly bad when the number of neighbors is very small (i.e., GNN has little information to aggregate). This is beacuse of the multi-layer in our model, it means that TMac can combine at least $l$ nodes for each node, thus ensuring that the model has basic neighborhood information to mine.

When $M_v=M_a=8$, TMac achieves the best performance. When the number of neighbors went up again, the performance went down instead. Due to the multiple layers setting of GNNs and the number of nodes in the constructed graph is actually relatively small, if the number of aggregated neighbors is high, it will lead to too similar representations of each node, which will bring unnecessary noise instead.

\subsubsection{Number of inter-modal neighbors}
Similar to the intra-modal hyper-parameter settings, we experimentally verified the number of inter-modal neighbors. From Fig. \ref{inter}, the fluctuation of the number of inter-modal neighbors is more obvious, and the effect is better at $M_c=5/8$, with the same ascending and then descending situation.

Although the performance fluctuates with the hyper-parameters, the overall effect on the experimental performance caused by the number of neighbors (both intra- and inter- modals), is maintained within a small range. This indicates that our proposed method TMac is not sensitive to the setting of hyper-parameters and is able to maintain a relatively stable performance.

\subsection{Convergence Analysis}
\textbf{Q4}: Will the extra consideration of time information affects the convergence of TMac? \textbf{Answer}: TMac can achieves convergence in a relatively short period.

\begin{figure}[t]
    \centering
    \includegraphics[width=0.35\textwidth]{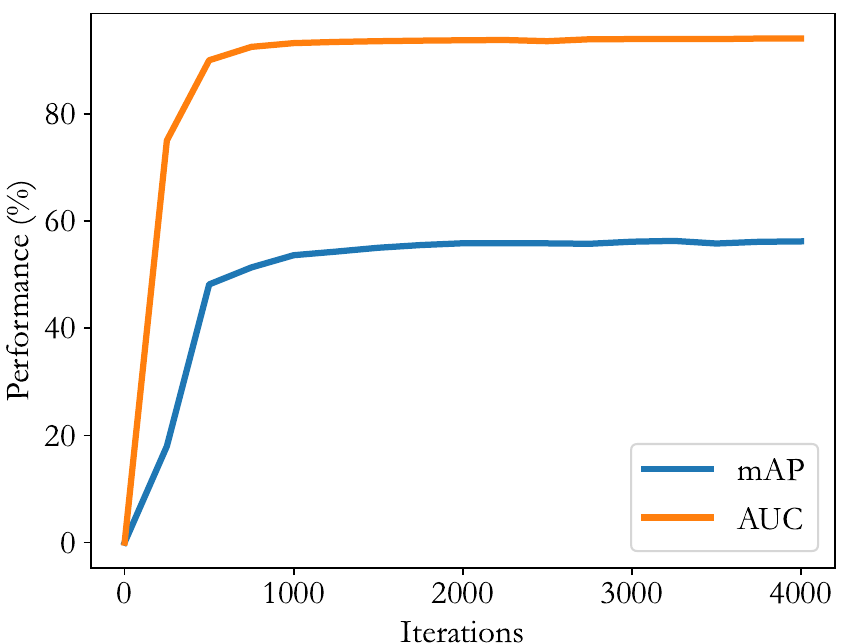}
    \caption{Performance changes over the number of iterations.}
    \label{conv}
\end{figure}

Here We discuss the convergence and running time of TMac. According Fig. \ref{conv}, we can find that the test performance of TMac increases rapidly, and begins to vibrate after iteration reaches 1000. As we implement one set of iterations per 250, and have five iterations to determine whether the iterations are most effective, there will be a long period of iterations. If the training speed is required, the model can achieve sub-optimal results just after 1000 iterations.

In addition, the time of each iteration is 0.4 seconds. Even considering the extra data reading time, the model can basically stop training in about 30-40 minutes, which is within the acceptable range. To sum up, after several groups of experiments, we can believe that TMac is meaningful for capturing time information.

\section{Conclusion}

In this paper, we point out the importance of temporal information in multi-modal acoustic event classification tasks. Due to the strong chronological order in video and audio data, it is natural to divide audiovisual data into multiple segments. Such segments can be consider as multi-modal nodes with different time relations for graph construction. We introduce a temporal graph neural network to capture both intra-modal and inter-modal dynamic information in the graph. Experimental results demonstrate the performance and effectiveness of TMac, and also show the importance of time information in audiovisual data modeling. In the future, we will explore more possibilities for cross-application between temporal graph learning and multi-modal scenarios.


\begin{acks}
This work was supported by the National Key R\&D Program of China (no. 2020AAA0107100), and the National Natural Science Foundation of China (no. 62325604, 62276271).
\end{acks}

\balance
\bibliographystyle{ACM-Reference-Format}
\bibliography{sample-base}


\end{document}